\documentclass[]{article}

\usepackage{color}
\usepackage[usenames,dvipsnames,svgnames,table]{xcolor}
\usepackage{graphicx}

\small

\newcommand{\msec}[2]{$#1\mbox{$''\mskip-7.6mu.\,$}#2$}

\begin{document}

\twocolumn

\begin{center}
\fboxrule0.02cm
\fboxsep0.4cm
\fcolorbox{Brown}{Ivory}{\rule[-0.9cm]{0.0cm}{1.8cm}{\parbox{7.8cm}
{ \begin{center}
{\Large\em Perspective}

\vspace{0.5cm}

{\Large\bf Distances to Star Forming Regions}

\vspace{0.2cm}

{\large\em Laurent Loinard}

% pls do not add affiliation

\vspace{0.5cm}

\centering
\includegraphics[width=0.21\textwidth]{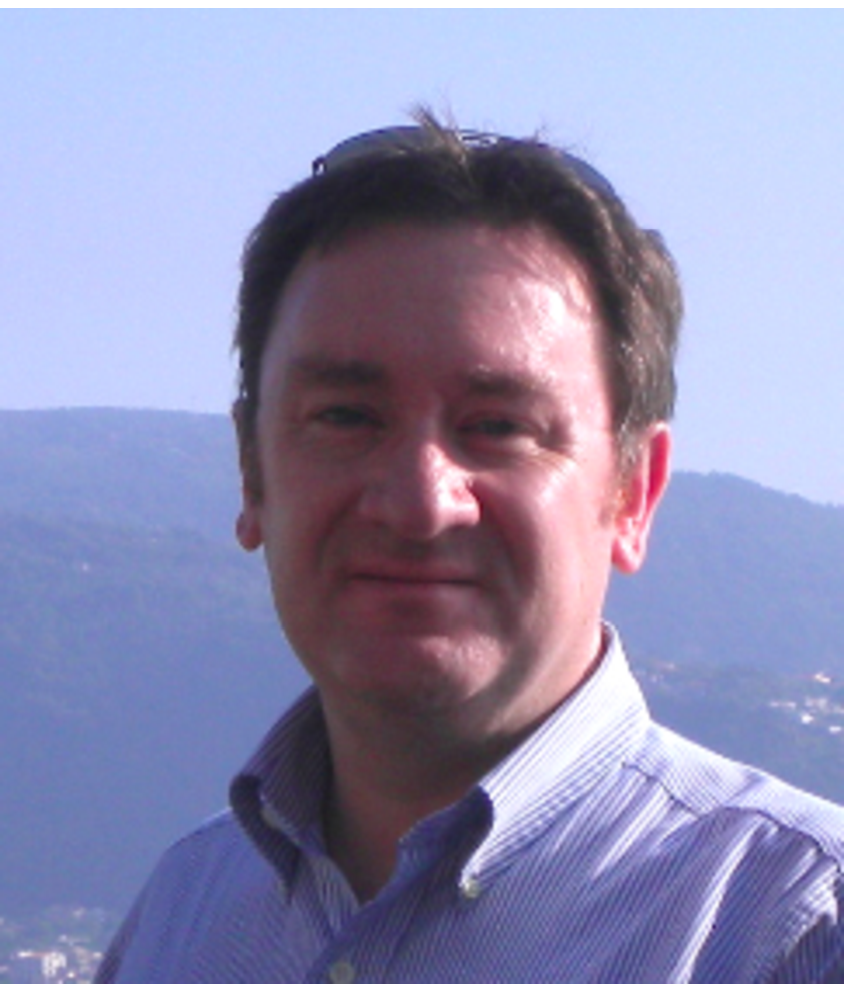}
\end{center}
}}}
\end{center}

\normalsize

\noindent
{\large \bf A historical perspective on astronomical \\ distance measurements}

The problem of accurate distance measurements has been a central theme of astronomy ever since the Antiquity (see de Grijs 2011 for a comprehensive contemporary discussion). For instance, Aristarchus of Samos is credited with the first reliable measurement of the distance to the Moon in the 4$^{th}$ century B.C.; remarkably, but somewhat fortuitously, his estimate was accurate to better than 1.5\% (e.g.\ Guillot 2001). Throughout the following centuries and up to this day, many ingenious methods have been devised to estimate distances to celestial bodies (a very important recent example is the determination of 
cosmological distances based on type Ia Supernovae; Riess et al.\ 1998, Perlmutter et al. 1999). It was clear from the beginning, however, that parallax techniques (i.e.\ those based on triangulation) were the most reliable, and this remains as true as ever.

The Merriam-Webster dictionary defines parallax as {\em ``the apparent displacement or the difference in apparent direction of an object as seen from two different points not on a straight line with the object.''} In principle, the two different points do not have to be very separated: Aficionados of the {\em Big Bang Theory} TV series (Season 1, Episode 1) may recall that Sheldon pick his ``spot'' in part because {\em ``it faces television at an angle that is (...) not so far wide as to create a parallax distortion''}. In the context of astronomy, parallaxes are used to estimate distances by measuring the angular difference in the apparent position on the celestial sphere of a given source observed from two different (widely separated, in this case) perspectives.\footnote{Thus, terms like {\em spectroscopic parallax} or {\em expansion parallax} are outright misnomers, since such methods involve no apparent displacement whatsoever. Their common use reflects the fact that astronomers have come to use ``parallax'' as a synonym for ``distance measurement method.'' } This was successfully realized for the Moon in 1751 by the French abbot and astronomer Nicolas Louis de La Caille (see Hirshfeld 2001) using simultaneous observations collected in Europe and the Cape of Good Hope (a baseline larger than 9,000 km). Similar observations were carried out in 1672 from Paris and Cayenne by Cassini and Richer to estimate the parallax of the planet Mars (see Hirshfeld 2001 and Ferguson 1999 for a discussion of the conclusiveness and accuracy of these results). At about the same time, Flamsteed obtained a similar value for Mars' parallax using a clever variant originally devised by Tycho Brahe. Rather than using two points on the Earth to define a long baseline, the same point can be used provided the observations are carried out several hours apart. Over such a time period, the rotation of the Earth on its axis carries the observer through space at a rate of about 1,000 km per hour (the displacement is of order of the Earth diameter for a timespan of 12 hours). A practical difficulty of this {\em diurnal parallax} method is that it requires fairly accurate time keeping, not easily achieved in the 16$^{th}$ and 17$^{th}$ centuries.

For objects outside of the Solar System, longer baselines are required to measure accurate parallaxes, and the solution comes from extending further Tycho's idea. Over the course of one year, the motion of the Earth about the Sun carries any point on the surface of the Earth through millions of kilometers. The resulting {\em trigonometric parallax} permits direct astronomical distance measurements provided the location of the Earth relative to the Sun is accurately known as a function of time. It is important to point out that at the level of accuracy reached by modern parallax measurements such as those that we will describe below, the requirement on the Earth-Sun separation can only be met by considering sophisticated Solar System dynamical models. Thus, although trigonometric parallaxes are commonly considered the first step of the cosmic distance ladder, they themselves rests on an accurate description of the Solar System dynamics. 

Returning to our historical development, we note that the first impetus to detect stellar trigonometric parallaxes came more from the desire to confirm the heliocentric Copernican view of the Universe, than from an interest about the distance to stars. The reasoning was that if the Earth were at the center of the Solar System (i.e.\ in the geocentric model of Ptolemy) distant stars would not exhibit any trigonometric parallax, whereas they would in the heliocentric Copernican model. The existence of stellar parallaxes thus became of prime importance to the development of astronomy during the Renaissance, and many astronomers (including Tycho, Galileo, and William Herschel) tried to measure them throughout the years. It was not until the 1830s, however,  that Friedrich Bessel, Thomas Henderson, and Friedrich von Struve successfully measured the trigonometric parallax of 61 Cygni, $\alpha$ Centauri, and Vega, respectively. The trigonometric parallaxes that they measured were only a fraction of an arcsecond (modern values are \msec{0}{29}, \msec{0}{75}, and \msec{0}{13} for 61 Cyg, $\alpha$ Cen, and Vega, respectively; van Leeuwen 2007), demonstrating that even the nearest stars are at considerable distances. This explained the lack of earlier detection, and implies that parallax measurements are indeed technically challenging: to measure the parallax of even the nearest star with 1\% accuracy requires angle measurements more accurate than 10 milli-arcseconds (mas)! 

Because of these technical requirements, less than 100 stars had trigonometric parallax measurements by the end of the 19$^{th}$ century, and many of them turned out to be very approximate. 
The situation steadily improved over the course of the 20$^{th}$ century, but the real breakthrough came with the Hipparcos mission (Perryman et al.\ 1997) of the European Space Agency (ESA), which measured the trigonometric parallax of over 100,000 stars with an average accuracy of about 1 mas. This translates into an accuracy on the distance better than 1\% for  stars within 10 pc and 10\% for stars within 100 pc. ESA has just launched a second astrometry mission (GAIA; de Bruijne 2012) that will provide trigonometric parallaxes for millions of stars. The full results are expected to be published in about 5 years, and the astrometric accuracy is anticipated to be of order 20 to 30 micro-arcseconds ($\mu$as) with some dependence on the stellar magnitude (de Bruijne 2012). This would result in distances accurate to better than 1\% for objects within 400 pc. 

\begin{figure*}[!t]
\centering
\includegraphics[width=0.85\textwidth]{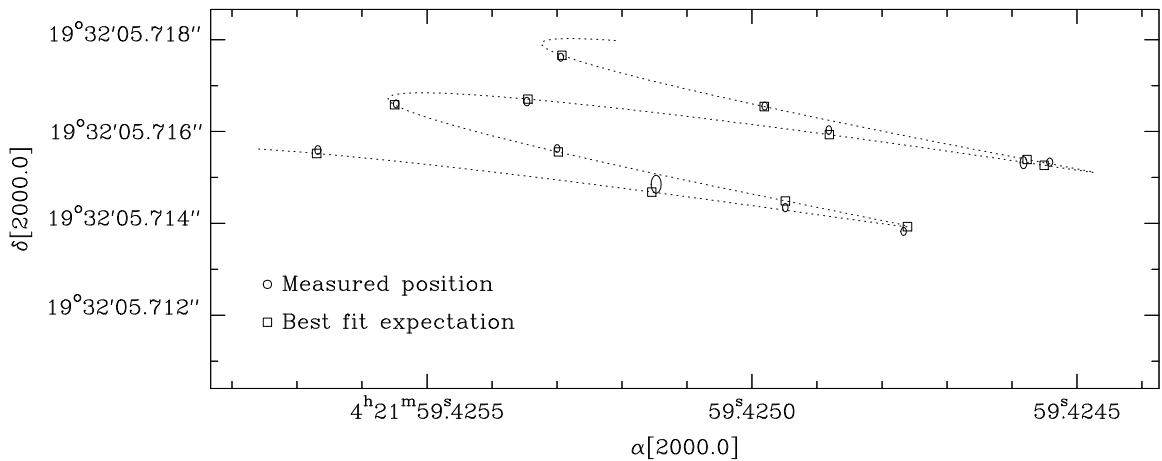}
\caption{Trajectory of T Tau Sb on the plane of the sky as measured by multi-epoch VLBA observations. The data were collected at 12 epochs from September 2003 to July 2005, and are shown as ellipses whose sizes represent the error bars. The dotted curve shows the best fit with a combination of parallax and proper motion. The squares are the positions predicted by the best fit at the twelve epochs (see Loinard et al. 2007 for details).}
\label{fig:fig1}
\end{figure*}

\bigskip

\noindent
{\large \bf Radio interferometry}

All the parallax measurements mentioned thus far were obtained from observations in the optical part of the electromagnetic spectrum. With the advent of radio-astronomy during the second half of the 20$^{th}$ century, and particularly with the development of radio-interferometry (Thompson et al.\ 2001), new astrometric tools became available. Indeed, it was rapidly clear that radio-interferometry was a powerful technique for astrometric measurement thanks to the relatively smaller effect of the Earth atmosphere on radio waves, and to the ease with which radio signals can be manipulated. The combination of these two factors enabled, in particular, the development of Very Long Baseline Interferometry (VLBI), a technique through which telescopes separated from one another by thousands of kilometers are operated as a single instrument\footnote{VLBI instruments can be constructed only because the signal can be adequately manipulated and recorded at each telescope, and sense can be made of the resulting interferences only because of the limited effect of the atmosphere.} to provide high angular resolution. At a wavelength of a few centimeters, the angular resolution achieved is of the order of a few mas. Concomitant with the high angular resolution of VLBI arrays comes their astrometric accuracy which --if all systematics can be controlled-- is given by $\theta$/SNR, where $\theta$ is the angular resolution and SNR the signal-to-noise with which the source is detected. Thus, with a signal-to-noise of a few hundred, an astrometric accuracy of order 10 $\mu$as can be achieved with VLBI arrays -- this is comparable with the accuracy expected of GAIA.

In principle, any set of radio telescopes operating at a common frequency can be tied up to form a VLBI array. For instance, the European VLBI Network (EVN; http://www.evlbi.org/intro/intro.html) operates by linking together up to two dozen radio telescopes distributed over Europe, Asia, and Africa. These telescopes are all different and are used most of the time as stand-alone ``single-dish'' telescopes; they are combined into a VLBI array only for a few weeks every year. This is very much like the situation with ESO's Very Large Telescope (VLT) where the Unit Telescopes (UT) are normally used by themselves, but can be linked with the Auxiliary Telescopes (AT) to form the VLTI. The National Radio Astronomy Observatory (NRAO) adopted a different strategy when it designed and built the Very Long Baseline Array (VLBA; http://www.vlba.nrao.edu) in the early 1990s. The VLBA is an array of 10 identical radio telescopes distributed over the US territory and operated solely as a VLBI array (i.e.\, they are never used as stand-alone telescopes). This has the great advantage of providing a VLBI array that can operate all year long (rather than only a few weeks every year), offering much greater scheduling flexibility. There is a common (and understandable) confusion outside of the radio community between VLBI and the VLBA. We hope it is now clear that VLBI is a technique, whereas the VLBA is a specific instrument using that technique: the VLBA is to VLBI what the VLTI is to optical interferometry.

\begin{figure*}[!t]
\centering
\includegraphics[width=0.85\textwidth]{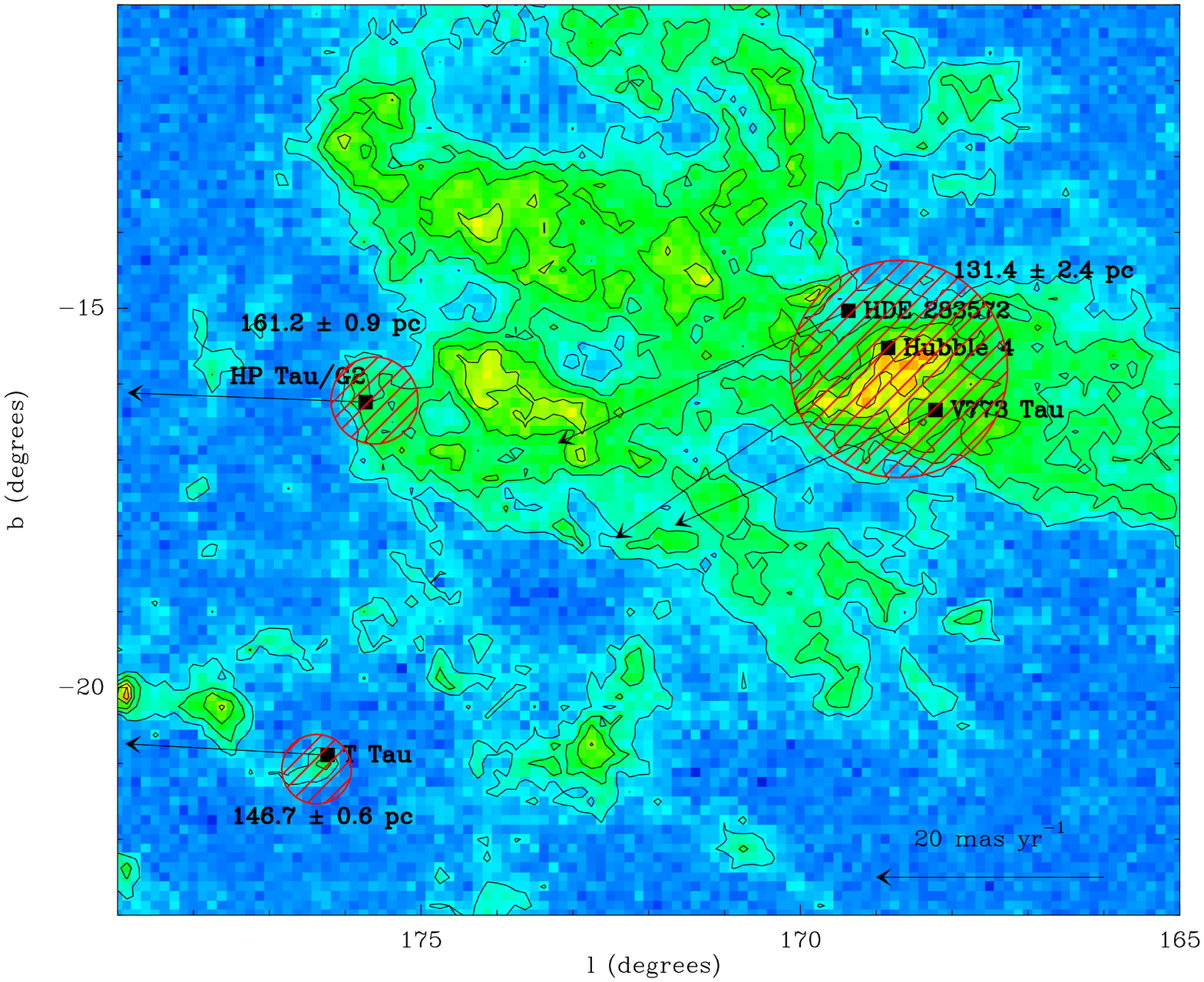}
\caption{Positions, distances, and proper motions of Hubble 4, HDE 283572, V773 Tau, T Tau, and HP Tau/G2 superposed on the CO(1-0) map of Taurus from Dame et al. (2001). Adapted from Loinard et al.\ (2007) and Torres et al.\ (2007, 2009, 2012).}
\label{fig:fig1}
\end{figure*}

The astrometric quality of VLBI instruments was used, in particular, to define the International Celestial Reference Frame (ICRF). The ICRF provides the closest approximation to date of an inertial reference frame and is materialized by precise equatorial coordinates for about 600 quasars distributed across the entire sky (Ma et al.\ 1998). VLBI arrays can also be used to measure trigonometric parallaxes to stars if they are sufficiently radio-bright. This was first achieved by Lestrade et al.\ (1999) who reached a typical accuracy of about 0.5 mas. Thanks to the improvement both in the sensitivity and in the calibration of VLBI observations over the past two decades, trigonometric parallaxes can now be measured with an accuracy of order 10 $\mu$as with VLBI arrays (Reid \& Honma 2014).

\begin{table*}
\caption{VLBI distances to star-forming regions within 1 kpc of the Sun.}
\medskip
\centering
\begin{tabular}{lll}
\hline \hline
Region & Distance & References\\%
\hline
Ophiuchus (L1688) & 120 $\pm$ 4 pc & Loinard et al.\ (2008) \\
Taurus-west       & 131.4 $\pm$ 2.4 pc       & Torres et al.\ (2007, 2012)\\
Taurus-east       & 161.2 $\pm$ 0.9 pc       & Torres et al.\ (2009)\\
Taurus-south     & 146.7 $\pm$ 0.6 pc       & Loinard et al.\ (2005, 2007)\\
Perseus (NGC 1333)     & 235 $\pm$ 18 pc & Hirota et al.\ (2008)\\
Perseus (L 1448)    & 232 $\pm$ 18 pc & Hirota et al.\ (2011)\\
Serpens core    & 415 $\pm$ 25 pc & Dzib et al.\ (2010)\\
Orion (ONC)        & 416 $\pm$ 5 pc & Menten et al.\ (2007); Kim et al.\ (2008)\\
Cepheus   & 700 $\pm$ 30 pc & Moscadelli et al.\ (2009); Dzib et al.\ (2011)\\
\hline \hline
\end{tabular}
\end{table*}

\bigskip

\noindent
{\large \bf Previous distances to star-forming regions}

Given the technical difficulty involved in obtaining accurate trigonometric parallaxes, distances to star-forming regions have traditionally been estimated from indirect techniques such as the convergent point method (e.g.\ Bertout \& Genova 2006; Galli et al.\ 2012) or ``spectroscopic parallaxes'' and isochrone fitting --both of which can be quite uncertain for young associations given the uncertainties of pre-main sequence evolutionary models and the substantial amount of dust that normally exists along the line of sight toward young stars. These indirect methods typically yield errors of 30\% or worse on the distance estimates of star-forming regions, and only provide a {\em mean distance} for each region. This is quite an unsatisfactory situation as a 30\% distance error propagates to uncertainties larger than 50\% on derived luminosities ($L \propto d^2$) and errors larger than 100\% on masses derived from Kepler's law ($M \propto a^3$). In addition, star-forming regions often exhibit internal structure, and in many cases, it is unclear whether a single mean distance is appropriate for the entire region, or if different sub-structures might be at different distances. 

Unfortunately, the Hipparcos satellite did not significantly improve our knowledge of the distances to star-forming regions (e.g.\ Bertout et al.\ 1999). This is because Hipparcos operated at optical wavelengths, and young stars are usually dim in the optical due to dust extinction. Indirect distances that incorporate Hipparcos results have enabled refinements in the distance determination of several star-forming regions (e.g.\ Knude \& H\o{}g 1998; Mamajek 2008), but with resulting uncertainties that remain substantial.

\bigskip

\noindent
{\large \bf VLBI distances to star-forming regions}

The situation started to change about 10 years ago thanks to VLBI measurements (see Loinard et al.\  2005). As we mentioned earlier, VLBI observations provide an astrometric accuracy similar to that expected from GAIA. Additionally, they have the distinct advantage that they are not affected by dust extinction. Thus, they can be used even for sources that are either deeply embedded within dusty regions, or located behind a large column of line-of-sight dust (or both). Star-forming regions are clearly in this situation, and are therefore prime targets for VLBI astrometric observations.

In star-forming regions, there are two classes of sources that are sufficiently bright at radio wavelengths to be detectable with VLBI arrays: masers and chromospherically active young stars. Masers (hydroxyl, water, or methanol) are commonly found in high-mass star-forming regions (Bartkiewicz \& van Langevelde 2012) and can be extremely bright. This makes them ideal tracers to map the distribution of high-mass star-forming regions across the Milky Way (Reid et al.\ 2014). Chromospherically active young stars, on the other hand, are often radio sources thanks to the gyration of electrons in their strong superficial magnetic fields (Dulk 1985). This results in continuum cyclotron, gyrosynchrotron, or synchrotron emission depending on the energy of the gyrating electrons. This emission is normally confined to regions extending only a few stellar radii around the stars, and therefore remains very compact even in the nearest star-forming regions (for instance, 4 $R_{\odot}$ $\equiv$ 50 $\mu$as at 300 pc). 

To appreciate the astrometric quality of VLBI observations, consider the case of T Tau Sb (Figure 1), one of the southern companions of the famous object T Tauri (e.g.\ Duch\^ene et al.\ 2002). T Tau Sb was observed with the Very Long Baseline Array (VLBA) at 12 epochs between 2003 September and 2005 July with a cadence of two months (see Loinard et al. 2007 for details). The position of the radio source associated with T Tau Sb was measured at each of these epochs and modeled as a combination of trigonometric parallax and proper motion. The best fit (shown as a dotted curve on Figure 1) yields a parallax of 6.82 $\pm$ 0.03 mas, corresponding to a distance $d$ = 146.7 $\pm$ 0.6 pc. Note that the resulting uncertainty on the distance is about 0.4\% which represents an improvement over the Hipparcos uncertainty of about two orders of magnitude.

Limiting ourselves to regions within 1 kpc of the Sun, similar observations have now been performed for a total of 5 young stars in Taurus (Torres et al.\ 2007, 2009, 2012), 2 young stars in Ophiuchus (Loinard et al.\ 2008), 2 young stars in Perseus (Hirota et al.\ 2008, 2011), 1 young star in Serpens (Dzib et al.\ 2010), 5 objects in the Orion Nebula Cluster (Hirota et al.\ 2007; Sandstrom et al.\ 2007; Menten et al.\ 2007; Kim et al.\ 2008), and 2 objects in Cepheus (Moscadelli et al.\ 2009; Dzib et al.\ 2011). The results are summarized in Table 1, and we consider them to be the most accurate distance estimates to date; note that we were careful to specify the sub-region in each case. Roughly 75\% of these distances are based on observations of the continuum emission emitted by chromospherically active young stars, whereas the other 25\% are from maser observations. In addition to these results for nearby regions ($d$ $<$ 1 kpc), roughly 100 parallaxes of similar accuracy (a few tens of $\mu$as) have been measured using VLBI observations of masers associated with high-mass star-forming regions distributed across the disk of the Milky Way (Reid et al.\ 2014; their Table 1). 

\bigskip

\noindent
{\large \bf VLBI tomography of star-forming regions}

As shown earlier, VLBI measurements can provide distances accurate to about 1 pc in the nearest star-forming regions. Such regions are typically a few tens of pc across, and are expected to be about as deep.
Thus, VLBI observations should easily resolve their depth, and could be used to perform {\em tomographic studies} of nearby regions. This has recently been demonstrated in the case of the Taurus star-forming region thanks to multi-source VLBA observations. As Figure 2 shows, the young stars located to the west of the Taurus region are at a common distance of about 130 pc, and share similar proper motions. Measurements from the literature show that they also have similar radial velocities. The stars to the east and south of the complex, on the other hand, are significantly farther (145-160 pc) and have different proper motions and radial velocities (from those to the west). Although based on a very limited number of targets, this example demonstrates that multi-source VLBI observations do enable {\it tomography} of individual star-forming regions. 

Such a possibility is fundamental to implement accurate astrophysics: Consider again the Taurus region, it appears to be at a mean distance of order 145 pc (the average of the near side at 130 pc and the far side at 160 pc), but to be 30 pc deep. Since 30 pc corresponds to 20\% of 145 pc, {\em using the mean distance indiscriminately for all stars in the complex results in a typical error of 20\%. } Accurate trigonometric parallaxes have eliminated any systematic errors on the mean distance, but statistical errors remain due to the intrinsic depth of the region. To eliminate them, the 3D structure of the region must be taken into account. We are currently in the process of measuring the trigonometric parallax of tens of young  stars distributed across the star-forming regions of Table 1 with the VLBA . This large project, called {\em The Gould's Belt Distances Survey} is described in detail in Loinard et al.\ (2011) and will start to provide results in the coming two years. Meanwhile, we encourage the community to use the best available distances not just for their favorite region, but also for their favorite sub-region. For instance, when considering stars on the western side of Taurus, 131.4 pc should be used rather than 140 or 145 pc. 

\bigskip

\noindent
{\large \bf Conclusions and prospects}

Ours are exciting times to consider star-forming region distances and structure. The VLBI results briefly described above have enabled a significant improvement in our best distance estimates and have shown that the internal structure of individual regions can be reconstructed. Interestingly, the internal kinematics can also be constrained by combining the high accuracy proper motions delivered by VLBI astrometry with radial velocity measurements. This type of results will be significantly expanded when the results of  {\em The Gould's Belt Distances Survey} currently underway at the VLBA are available. 

Of course, GAIA will soon provide results with an accuracy similar to those delivered by VLBI observations.The two sets of data will be extremely complementary. First they will be useful to discard any systematic effects in either data sets (see Melis et al.\ 2014 for a discussion of the ``Pleiades Controversy'', a case where Hipparcos and VLBI results are in disagreement). In addition, their combination will be a powerful probe of the structure of nearby regions: GAIA will deliver a large number of parallaxes but will be limited to regions of moderate extinction. This could induce systematic errors: If we assume that a typical star-forming region contains clumps with an $A_V$ $\sim$ 10, then GAIA will be biased to stars located in front (or at least on the near side) of such clumps. VLBI experiments are not affected by dust extinction and will identify regions where GAIA results might be biased and correct the distances appropriately.

Two final points are worth mentioning. First, that even higher astrometric accuracy (5 $\mu$as or better) should be achievable through VLBI experiments incorporating the Square Kilometer Array (SKA; Loinard et al.\ 2015). This would correspond to a breathtaking 0.06\% accuracy (less than a tenth of a pc) on individual distances in Ophiuchus. SKA-VLBI experiments would also enable tomography mappings similar to those shown here for Taurus for regions at a few kpc. Second, that for highly obscured regions, a GAIA-like mission with near-infrared detectors might be useful. The situation is not so clear, however, as discussed by H\o{}g \& Knude (2014).

\bigskip

\footnotesize

{\bf References:}\\

Bartkiewicz, A., \& van Langevelde, H.~J.\ 2012, IAU Symposium, 287, 117 \\
Bertout, C., Genova, F., A\&A, 460, 499, 2006\\
Bertout, C., Robichon, N., Arenou, F., A\&A, 352, 574, 1999\\
de Bruijne, J.H.J.\ 2012, Astrophysics and Space Science, 341, 31 \\
Duchêne, G.; Ghez, A.M.; McCabe, C., ApJ, 568, 771, 2002\\
Dulk, G.~A.\ 1985, ARAA, 23, 169 \\
Dzib, S., Loinard, L., Mioduszewski, A.J., et al., ApJ, 718, 610, 2010\\
Dzib, S., Loinard, L., Rodríguez, L.F., et al., ApJ, 733, 71, 2011\\
Ferguson, K., ``Measuring the Universe'', Walker \& Company. 1999\\
Galli, P.A.B., Teixeira, R., Ducourant, C., et al., A\&A, 538, 23, 2012\\
Gaume, R.A., Wilson, T.L., Vrba, F.J., Johnston, K.J., \& Schmid-Burgk, J., 1998, ApJ, 493, 940 \\
de Grijs, R., ``An Introduction to distance measurements in astronomy'', Wiley 2011\\
Guillot, D.J., $http://www.bgfax.com/school/distance\_history.pdf$, 2001\\
Hirota, T., Honma, M., Imai, H., et al., PASJ, 63, 1, 2011\\
Hirota, T., Bushimata, T., Choi, Y.-K., et al., PASJ, 60, 37, 2008\\
Hirota, T., Bushimata, T., Choi, Y.-K., et al., PASJ, 59, 897, 2007\\
Hirschfeld, A.,  ``Parallax: The Race to Measure the Cosmos'', W.H. Freeman and Company, 2001\\
H\o{}g E., Knude, J., arXiv:1408.3305, 2014\\
Kim, M.~K., Hirota, T., Honma, M., et al., PASJ, 60, 991, 2008\\
Knude, J., \& H\o{}g, E., A\&A, 338, 897, 1998\\
van Leeuwen, F., ``Hipparcos: the new reduction of the raw data'', Astrophysics and Space Science Library, Springer, 2007\\
Lestrade, J.-F., Preston, R.A., Jones, D.L., et al., A\&A, 344, 1014, 1999\\
Loinard, L., Mioduszewski, A.J., Torres, R.M., RMxAC, 40, 205, 2011\\
Loinard, L., Thompson, M., Hoare, M., et al., ``Science with the SKA'', 2015, in prep.\\
Loinard, L., Torres, R.M., Mioduszewski, A.J., et al., ApJ, 671, 546, 2007\\
Loinard, L., Torres, R.M., Mioduszewski, A.J., \& Rodr{\'{\i}}guez, L.F., ApJ, 675, L29, 2008\\
Ma, C., Arias, E.F., Eubanks, T.M., et al., AJ, 116, 516, 1998\\
Mamajek, E. E., Astronomische Nachrichten, 322, 10, 2008\\
Melis, C., Reid, M.J., Mioduszewski, A.J., et al., arXiv:1408.6544, 2014\\
Menten, K.M., Reid, M.J., Forbrich, J., \& Brunthaler, A., ApJ, 474, 515, 2007\\
Moscadelli, L., Reid, M.J., Menten, K.M., et al., ApJ,  693, 406, 2009\\
Perlmutter, S., Aldering, G., Goldhaber, G, et al., ApJ, 517, 565, 1999\\
Perryman, M.A.C., Lindegren, L., Kovalevsky, J., et al., A\&A, 323, 49, 1997\\
Reid, M.J., Honma, M., ARAA, 52, 339, 2014\\
Reid, M.J., Menten, K. M., Brunthaler, A., et al., ApJ, 783, 130, 2014\\
Riess, A.G., Filippenko, A.V., Challis, P., et al., AJ, 116, 1009, 1998\\
Sandstrom, K.M., Peek, J.E.~G., Bower, G.C., et al., ApJ, 667, 1161, 2007\\
Thompson, A.R., Moran, J.M., \& Swenson, G.W., Interferometry and Synthesis in Radio Astronomy, John Wiley \& Sons, 2007.\\
Torres, R.M., Loinard, L., Mioduszewski, A.J., et al., ApJ, 747, 18, 2012\\
Torres, R.M., Loinard, L., Mioduszewski, A.J., \& Rodr{\'{\i}}guez, L.F., ApJ, 698, 242, 2007\\
Torres, R.M., Loinard, L., Mioduszewski, A.J., \& Rodr{\'{\i}}guez, L.F., ApJ, 671, 1813, 2007\\

\end{document}